\documentclass[english,11pt,nofootinbib,showpacs,notitlepage]{revtex4-1}
\usepackage[T1]{fontenc}
\usepackage{babel}
\usepackage[utf8]{inputenc}
\usepackage{amssymb,mathrsfs,amsfonts,amsmath,euscript,textcomp,graphicx,wrapfig,color,txfonts,lmodern,multirow,dsfont}
\usepackage[section]{placeins}
\usepackage[usenames,dvipsnames,svgnames,table]{xcolor}
\definecolor{darkgreen}{rgb}{0,.7,0}
\usepackage[colorlinks,linkcolor=blue,citecolor=darkgreen,pagebackref=true]{hyperref}
\def\dac{\displaystyle\frac}

\def\[{\left[}
\def\]{\right]}
\def\({\left(}
\def\){\right)}

\newcommand{\eq}[1]{\begin{equation}#1\end{equation}}
\newcommand{\eqs}[1]{\begin{equation*}#1\end{equation*}}
\newcommand{\oneN}{=\overline{1,N}}

\newcommand{\const}{\mathop{\rm const}\nolimits}

\newcommand{\e}{\mathop{\rm e}\nolimits}

\newcommand{\bgbr}[1]{\bigl\{#1\bigr\}}

\newcommand{\bgsb}[1]{\bigl[#1\bigr]}
\newcommand{\Bgp}[1]{\Bigl(#1\Bigr)}
\newcommand{\bgp}[1]{\bigl(#1\bigr)}

\newcommand{\nd}{\noindent}

\begin{document}
\baselineskip7mm

\title{Constant volume exponential solutions in Einstein-Gauss-Bonnet flat anisotropic cosmology with a perfect fluid}

\author{Dmitry Chirkov}
\affiliation{Sternberg Astronomical Institute, Moscow State University, Moscow 119991 Russia}
\affiliation{Faculty of Physics, Moscow State University, Moscow 119991 Russia}
\author{Sergey A. Pavluchenko}
\affiliation{Instituto de Cinesis F\'isicas y Matem\'aticas, Universitas Austral de Chile, Valencia, Chile}
\author{Alexey Toporensky}
\affiliation{Sternberg Astronomical Institute, Moscow State University, Moscow 119991 Russia}

\begin{abstract}
In this paper we investigate the constant volume exponential solutions (i.e. the solutions with the scale factors change exponentially over time so that the
comoving volume remains the same) in the Einstein-Gauss-Bonnet gravity.
We find conditions for these solutions to exist and show that they are compatible with any perfect fluid with the equation of state
parameter $\omega<1/3$ if the matter density of the Universe exceeds some critical value.
We write down some exact solutions which generalize ones found in our previous paper for models with a cosmological constant.


\end{abstract}



\maketitle

\section{Introduction}

Exact solutions play important role in any gravitational theory, especially nonlinear. Indeed, using numerical recipes
one almost always can build a solution, but its viability will be questioned. This is especially true for nonlinear theories where even numerical
solutions are sometimes hard to find.

Lovelock gravity~\cite{Lovelock} is the striking example of the nonlinear theory of gravity. It is the most general metric theory of gravity yielding
conserved second order equations of motion (in contrast to $f(R)$ gravity which gives fourth order dynamical equations) in arbitrary number of spacetime
dimensions. One can say that the Lovelock gravity is a natural generalization of Einstein's General Relativity in the following sense: it is
known~\cite{etensor1, etensor2, etensor3} that the Einstein tensor is, in any dimension, the only symmetric and conserved tensor depending only on the
metric and its first and second derivatives (with a linear dependence on second derivatives); if one drops the condition of linear dependence on second
derivatives, one can obtain the most general tensor which satisfies other mentioned conditions -- the Lovelock tensor.

The Lovelock gravity has been intensively studied in the cosmological context (see, e.g.,\cite{add_1, add_2, add_3, add_4, add_6, add_7,
add_8, add_10, add13, add_11, mpla09, prd10}).
Particularly, many interesting results have been obtained for flat anisotropic metrics due to the fact that its cosmological dynamics is much richer in the
Lovelock gravity than in the Einstein one. Since the resulting equations of motion turn out to be complicated enough, researchers usually study some special
kind of metric (e.g. with only two different scale factors~\cite{add13, CGP}) or consider Lagrangian that contain the highest order Lovelock term only
(e.g., deleting Einstein term and keeping Gauss-Bonnet term in a Lagrangian in the cases of (4+1)- and (5+1)-dimensional spacetimes one get so called
``pure'' Gauss-Bonnet model -- see, for instance, \cite{Pavluchenko, grg10, Ivashchuk}). In the latter approach solutions with power-law and exponential time dependence
of scale factors were found. The first of them
is an analog of Kasner solution~\cite{Deruelle1,Deruelle2} -- scale factors in this solution have power-law behavior, though relations between power indices
is different from the Kasner solution in Einstein gravity~\cite{add_12, Pavluchenko, PT}. Special features of the second type of  solutions -- Hubble parameters are constant, so
in a flat metric differential equations of motion become algebraic -- allows us to study them  in more complicated theories~\cite{KPT, PT}.

There is a meaning behind considering these two metric {\it ansatz} -- power-law and exponential -- while looking for exact solutions. The former of them could be
considered as a ``classical'' Friedman power-law expansion, but generalized for flat anisotropic metric. So that finding generalized Kasner power-law solutions we find
possible ``Friedman-like'' attractors in high-curvature regime for the general system.
The latter could be considered as anisotropic generalization of the de Sitter (exponential) expansion. Unlike power-law solutions which could be build only when one
(usually highest) Lovelock term is considered, exponential solutions could be obtained when a mixture of Lovelock terms is considered. It makes exponential solutions
more related to general Lovelock theory then power-law ones; from physical point of view they could be considered as ``inflation-like'' attractors\footnote{
But the analogy is not totally correct -- indeed, generally by ``inflation'' is meant not any exponential expansion stage, but the one with a mechanism to end this stage,
from this point of view exponential solution cannot be called ``inflation'' for it will never ends.}. For this reason we are looking for exact solutions of this kind.
When considering general cases, solutions of this kind also often are found (see e.g.~\cite{add13, CGP, MO14}; in particular, in~\cite{CGP} solutions of this kind were
found in the model with non-flat spatial sections), and as one of the goals we want to describe the general conditions for these solutions to exist.

In our previous paper~\cite{CST} we started to investigate the exponential solutions in Einstein-Gauss-Bonnet gravity. In the course of the study we have
shown that these solution are divided into two different types -- with constant volume and with volume changing in time. The paper~\cite{CST} is devoted to
the latter case. In the present paper we consider solutions with constant volume.

The structure of the manuscript is as follows: in the second section we introduce the set-up we are working on and very briefly reintroduce the results from
our previous paper. Then in Section III we write down solutions of a special structure which generalize
those found in~\cite{CST} and in Section IV finally we work with a general case. Section V
concludes the results of this paper and compares them with results of our previous paper~\cite{CST}.

\section{The set-up.}
\nd The Einstein-Gauss-Bonnet action in $(N+1)$-dimensional spacetime reads\footnote{Throughout the paper we use the system of units in which $c=1$, $c$ is the speed of light. Greek indices run from 0 to N, Latin indices from 1 to N.}:
\eq{S=\frac{1}{2\kappa^2}\int d^{N+1}x\sqrt{-g}\bgp{\mathcal{L}_E+\alpha\mathcal{L}_{GB}+\mathcal{L}_m},\quad\mathcal{L}_E=R,\quad
\mathcal{L}_{GB}=R_{\alpha\beta\gamma\delta}R^{\alpha\beta\gamma\delta}-4R_{\alpha\beta}R^{\alpha\beta}+R^2,\label{action}}
where $\kappa^2$ is the $(N+1)$-dimensional gravitational constant, $R,R_{\alpha\beta},R_{\alpha\beta\gamma\delta}$ are the $(N+1)$-dimensional scalar curvature, Ricci tensor and Riemann tensor respectively, $\alpha$ is the coupling constant $\mathcal{L}_m$ is the Lagrangian of a matter. The action~(\ref{action}) gives the gravitational equations as
\eq{G_{\mu\nu}+\alpha H_{\mu\nu}=\kappa^2T_{\mu\nu}\label{dynamical.eqs.general.form}}
where
\eq{G_{\mu\nu}=R_{\mu\nu}-\frac{1}{2}g_{\mu\nu}R,\quad T_{\mu\nu}=-2\,\frac{\delta\mathcal{L}_m}{\delta g_{\mu\nu}}+g_{\mu\nu}\mathcal{L}_m\label{G_munu.T_munu}}
\eq{H_{\mu\nu}=2\(RR_{\mu\nu}-2R_{\mu\alpha}R^{\alpha}_{\;\nu}-2R^{\alpha\beta}R_{\mu\alpha\nu\beta}+R_{\mu}^{\;\alpha\beta\gamma}R_{\nu\alpha\beta\gamma}
-\frac{1}{2}g_{\mu\nu}\mathcal{L}_{GB}\)}
We consider a perfect fluid with the equation of state $p=\omega\rho$ as a matter source. The spacetime metric is
\eq{ds^2=-dt^2+\sum_{k=1}^N\e^{2H_kt}dx_k^2,\quad H_k\equiv\const\label{metric}}
It is easily shown that
\eq{R^{0i}_{0i}=H_i^2,\quad R^{j_1j_2}_{j_1j_2}=H_{j_1}H_{j_2},\;\;j_1<j_2,\quad R^{\alpha\beta}_{\mu\nu}=0,\;\;\bgbr{\alpha,\beta}\ne\bgbr{\mu,\nu},\label{components.of.the.Riemann.tensor}}
the dot denotes derivative w.r.t. $t$. So, arbitrary component of the Riemann tensor takes the form:
\eq{R^{\mu\nu}_{\lambda\sigma}=
\left\{\sum\limits_{k}H_k^2\delta_0^{[\mu}\delta_k^{\nu]}\delta^0_{[\lambda}\delta^k_{\sigma]}+
\sum\limits_{i<j}H_iH_j\delta_i^{[\mu}\delta_j^{\nu]}\delta^i_{[\lambda}\delta^j_{\sigma]}\right\},\label{gen.expr.for.Riemann}}
square brackets denote the antisymmetric part on the indicated indices.
In view of~(\ref{G_munu.T_munu})-(\ref{gen.expr.for.Riemann}) equations~(\ref{dynamical.eqs.general.form}) take the form
\eq{2\sum\limits_{i\ne j}H_i^2+2\sum\limits_{\{i>k\}\ne j}H_i H_k+8\alpha\sum\limits_{i\ne j}H_i^2\sum\limits_{\{k>l\}\ne\{i, j\}}H_k H_l+24\alpha\sum\limits_{\{i>k>l>m\}\ne j}H_i H_k H_l H_m=-\omega\varkappa,\;\;j\oneN\label{eq.of.motion}}
\eq{2\sum\limits_{i>j}H_i H_j+24\alpha\sum\limits_{i>j>k>l}H_i H_j H_k H_l=\varkappa,\quad\varkappa=\kappa^2\rho\label{constraint}}
Continuity equation reads:
\eq{\dot\rho+(\rho+p)\sum_i H_i=0\label{continuity.eq}}
Subtracting $i$-th dynamical equation from $j$-th one we obtain:
\eq{\(H_j - H_i\)\(\frac{1}{4\alpha}+\sum\limits_{\{k>l\}\ne\{ i, j\}}H_k H_l\)\sum_k H_k=0\iff
\left[\begin{array}{r}
        H_{i}=H_{j}\quad\mathbf{(i)} \vspace{.2cm}\\
        \sum\limits_{\{k>l\} \ne \{ i, j\}} H_k H_l=-\frac{1}{4\alpha}\quad\mathbf{(ii)} \vspace{.2cm}\\
        \sum_k H_k=0\quad\mathbf{(iii)}
      \end{array}\right.
\label{after.subtraction}}
Left hand sides of Eqs.~(\ref{eq.of.motion})--(\ref{constraint}) does not depend on time, therefore $\rho\equiv\const$, so that the Eq.~(\ref{continuity.eq})
reduces to
\eq{\label{rho_all}(\rho+p)\sum_i H_i=0\iff
\left[\begin{array}{r}
        \rho=0\quad\mathbf{(a)} \vspace{.2cm}\\
        p=-\rho\quad\mathbf{(b)} \vspace{.2cm}\\
        \sum_k H_k=0\quad\mathbf{(c)}
      \end{array}\right.}
For a given set $H_1,\ldots,H_N$ to be a solution of Eqs.~(\ref{eq.of.motion})--(\ref{constraint})
it is necessary that at least one of the conditions \textbf{(i)-(iii)} is satisfied.
In our previous work~\cite{CST} we considered situations when conditions \textbf{(i),(ii)} and their combinations are satisfied; it was found that taking into account these conditions lead to consistent system of equations for the vacuum (\textbf{a}) and $\Lambda$-term (\textbf{b}) cases only. In the present manuscript we interest in the condition \textbf{(iii)} and its combination with conditions \textbf{(i),(ii)}. Requirement $\sum_k H_k=0$ does not impose any constraints on choice of a matter from the continuity equation; we will see later that
additional constraints on the form of the perfect fluid is  followed from equations for the gravitational field.
\section{Constant volume solutions with two different Hubble parameters.}
In the present section we generalize solutions found in~\cite{CST} to an arbitrary equation of state of the matter. Taking into account results
of the cited paper we assume that there are only two different parameters in the set $H_1,\ldots,H_N$. Now one can easily obtain a number of special exact constant volume solutions. We consider three basic cases which are of great importance for low-dimensional spacetimes with $N=4,5$.

Using condition $\sum_i H_i=0$ it is easy to check that
\eq{2\sum\limits_{j_1<j_{2}}H_{j_1}H_{j_2}=-\sum\limits_{i}H_i^2\label{H1H2}}
\eqs{24\sum\limits_{j_1<\ldots<j_{4}}H_{j_1}H_{j_2}H_{j_3}H_{j_4}=
-\sum\limits_{i}H_i^4-4\sum\limits_{j_1}H_{j_1}^3\sum\limits_{j_{2}\ne j_1}H_{j_2}-3\sum\limits_{j_1}H_{j_1}^2\sum\limits_{j_{2}\ne j_1}H_{j_2}^2-12\sum\limits_{j_1}H_{j_1}^2\sum\limits_{j_{2}<j_3\atop i_2,j_3\ne j_1}H_{j_2}H_{j_3}=}
\eqs{=-\sum\limits_{i}H_i^4+4\sum\limits_{j_1}H_{j_1}^4-
3\sum\limits_{j_1}H_{j_1}^2\sum\limits_{j_{2}\ne j_1}H_{j_2}^2-
6\sum\limits_{j_1}H_{j_1}^2\left[H_{j_1}^2-\sum\limits_{j_2\ne j_1}H_{j_2}^2\right]=}
\eq{=-3\sum\limits_{i}H_i^4+3\sum\limits_{j_1}H_{j_1}^2\sum\limits_{j_{2}\ne j_1}H_{j_2}^2=3\left[\sum\limits_{j_1}H_{j_1}^2\right]^2-6\sum\limits_{i}H_i^4\label{H1H2H3H4}}
Substituting~(\ref{H1H2}) and~(\ref{H1H2H3H4}) into the Eqs.~(\ref{eq.of.motion})--(\ref{constraint}) we obtain:
\eq{\left\{\begin{array}{c}
             -\sum\limits_{i}H_i^2+\alpha\left(\left[\sum\limits_{j_1}H_{j_1}^2\right]^2-2\sum\limits_{i}H_i^4\right)=\omega\varkappa \\
             \sum\limits_{i}H_i^2-3\alpha\left(\left[\sum\limits_{j_1}H_{j_1}^2\right]^2-2\sum\limits_{i}H_i^4\right)=-\varkappa
           \end{array}
\right.\label{final.part.sys}}
One can see that in the vacuum case ($\varkappa=0$) the system~(\ref{final.part.sys}) has no nontrivial\footnote{We call a solution trivial if $H_1=\ldots=H_N=0$.} solution, except for the situation of pure Gauss-Bonnet model (the first term in both Eqs.~(\ref{final.part.sys}) is absent) -- the corresponding solution was found by~\cite{Ivashchuk}. It naturally follows from~(\ref{final.part.sys}) that
\eq{\sum\limits_{i}H_i^2=-3\(\omega-\dac{1}{3}\)\(\dac{\varkappa}{2}\),\quad \sum\limits_{i}H_i^4=\frac{1}{2}\left[9\(\omega-\dac{1}{3}\)^2\(\dac{\varkappa}{2}\)^2+\dac{\omega-1}{\alpha}\(\dac{\varkappa}{2}\)\right]\label{h2h4}}
\begin{enumerate}
  \item $\bgp{[N-1]+1}$-decomposition: $H_1=\ldots=H_{N-1}\equiv H\in\mathbb{R}$, $H_N \equiv h\in\mathbb{R}$. It follows from the condition $\sum_i H_i = 0$ that $h = -(N-1)H$. Substituting these $H_1,\ldots,H_N$ into Eqs.~(\ref{h2h4}) we obtain:
      \eq{\left\{\begin{array}{c}
                 N(N-1)H^2=-3\(\omega-\dac{1}{3}\)\(\dac{\varkappa}{2}\)\vspace{.3cm}\\
                 2N(N-1)(N^2-3N+3)H^4=9\(\omega-\dac{1}{3}\)^2\(\dac{\varkappa}{2}\)^2+\dac{\omega-1}{\alpha}\(\dac{\varkappa}{2}\)
               \end{array}\right.
    \label{N-1+1}}
    Solution of Eqs.~(\ref{N-1+1}) for $H^2$ and $\rho$:
    \eq{H^2=-\frac{1}{3\alpha(N-2)(N-3)}\,\frac{\omega-1}{\omega-\frac{1}{3}},\quad\rho=\frac{1}{36\pi\alpha}\,\frac{N(N-1)}{(N-2)(N-3)}\,
    \frac{\omega-1}{\(\omega-\frac{1}{3}\)^2},\quad\omega<\frac{1}{3},\quad\alpha<0\label{sol.for.N-1+1}}
  \item $\(\frac{N}{2}+\frac{N}{2}\)$-decomposition, $N$ is even: $H_1=\ldots=H_{\frac{N}{2}}\equiv H\in\mathbb{R}$, $H_{\frac{N}{2}+1}=\ldots=H_N \equiv h\in\mathbb{R}$. It follows from the condition $\sum_i H_i = 0$ that $h = -H$. Substituting these $H_1,\ldots,H_N$ into Eqs.~(\ref{h2h4}) we obtain:
      \eq{\left\{\begin{array}{c}
                 NH^2=-3\(\omega-\dac{1}{3}\)\(\dac{\varkappa}{2}\)\vspace{.3cm}\\
                 2NH^4=9\(\omega-\dac{1}{3}\)^2\(\dac{\varkappa}{2}\)^2+\dac{\omega-1}{\alpha}\(\dac{\varkappa}{2}\)
               \end{array}\right.
    \label{N/2+N/2}}
    Solution of Eqs.~(\ref{N/2+N/2}) for $H^2$ and $\rho$:
    \eq{H^2=\frac{1}{3\alpha(N-2)}\,\frac{\omega-1}{\omega-\frac{1}{3}},\quad\rho=-\frac{1}{36\pi\alpha}\,\frac{N}{N-2}\,
    \frac{\omega-1}{\(\omega-\frac{1}{3}\)^2},\quad\omega<\frac{1}{3},\quad\alpha>0\label{sol.for.N/2+N/2}}
  \item $\([n+1]+n\)$-decomposition, $n\equiv\left\lfloor\frac{N}{2}\right\rfloor$, $N$ is odd: $H_1=\ldots=H_{n+1}\equiv H\in\mathbb{R}$, $H_{n+2}=\ldots=H_N \equiv h\in\mathbb{R}$. It follows from the condition $\sum_i H_i = 0$ that $h = -(1+n^{-1})H$. Substituting these $H_1,\ldots,H_N$ into Eqs.~(\ref{h2h4}) we obtain:
      \eq{\left\{\begin{array}{c}
                 N(1+n^{-1})H^2=-3\(\omega-\dac{1}{3}\)\(\dac{\varkappa}{2}\)\vspace{.3cm}\\
                 2N(1+n^{-1})(1+n^{-1}+n^{-2})H^4=9\(\omega-\dac{1}{3}\)^2\(\dac{\varkappa}{2}\)^2+\dac{\omega-1}{\alpha}\(\dac{\varkappa}{2}\)
               \end{array}\right.
    \label{N/2+1+N/2}}
    Taking into account $n=\frac{N-1}{2}$ we get solution of Eqs.~(\ref{N/2+N/2}) for $H^2$ and $\rho$:
    \eq{\begin{array}{c}
          H^2=\dac{1}{3\alpha}\,\dac{(N-1)^2}{(N-3)(N^2+N+2)}\,\dac{\omega-1}{\omega-\frac{1}{3}},\quad
    \rho=-\dac{1}{36\pi\alpha}\,\dac{N(N-1)(N+1)}{(N-3)(N^2+N+2)}\,\dac{\omega-1}{\(\omega-\frac{1}{3}\)^2}, \\
          \omega<\dac{1}{3},\quad\alpha>0
        \end{array}\label{sol.for.N/2+1+N/2}}
\end{enumerate}
For $N=4$ only cases 1 and 2 are realized: it is (3+1)-decomposition and (2+2)-decomposition; for $N=5$ only cases 1 and 3 are realized: it is (4+1)-decomposition and (3+2)-decomposition and there are no other options for $N=4,5$. It is easy to check that for $N=4,5$ and $\omega=-1$ solutions~(\ref{sol.for.N-1+1}),(\ref{sol.for.N/2+N/2}),(\ref{sol.for.N/2+1+N/2}) turn to solutions derived in our previous paper~\cite{CST} with additionally imposed constant volume requirement $\sum_i H_i = 0$.

For $N=5$ there is one more decomposition, containing 3 different Hubble parameters (see~\cite{CST}): $H_1=H_2\equiv H,\;H_3=H_4\equiv -H,\;H_5\equiv h$; but it follows from $\sum_i H_i = 0$ that $h=0$ and it reduces to $(2+2)$-decomposition; for the same reasons decomposition $H_1=\ldots=H_{n}\equiv H\in\mathbb{R}$, $H_{n+1}=\ldots=H_{2n} \equiv -H$, $H_N\equiv h\in\mathbb{R}$ $\(n\equiv\left\lfloor\frac{N}{2}\right\rfloor,\;N\;\mbox{is odd}\)$ reduces to $(n+n)$-decomposition described above. So, all
possible generalization of solutions in $(4+1)$ and $(5+1)$ dimensions
found in \cite{CST} for $w=-1$ to an arbitrary $w$ are presented in our list.

We should note that in a general set-up (see the next section) other decompositions (for example, $(2+1+1)$ in $(4+1)$ dimensions)
are possible and can be found by inserting corresponding ansatz into Eqs.~(\ref{final.part.sys}). However, such solutions represent special cases of
general solution with constant volume, and, unlike written down above, have no connections with varying volume solutions found
in \cite{CST}.

\section{Necessary conditions for general constant volume solutions.\label{CVE}}
In general case of constant volume solution we do not expect any additional relations between Hubble parameters (in contrast to
the varying volume case, where only space-times with isotropic subspaces are possible). The full set of solution is rather cumbersome
to be written down explicitly, so we restrict ourselves by finding conditions of its existence.

\begin{figure}[!h]
\center{\includegraphics[width=.9\linewidth]{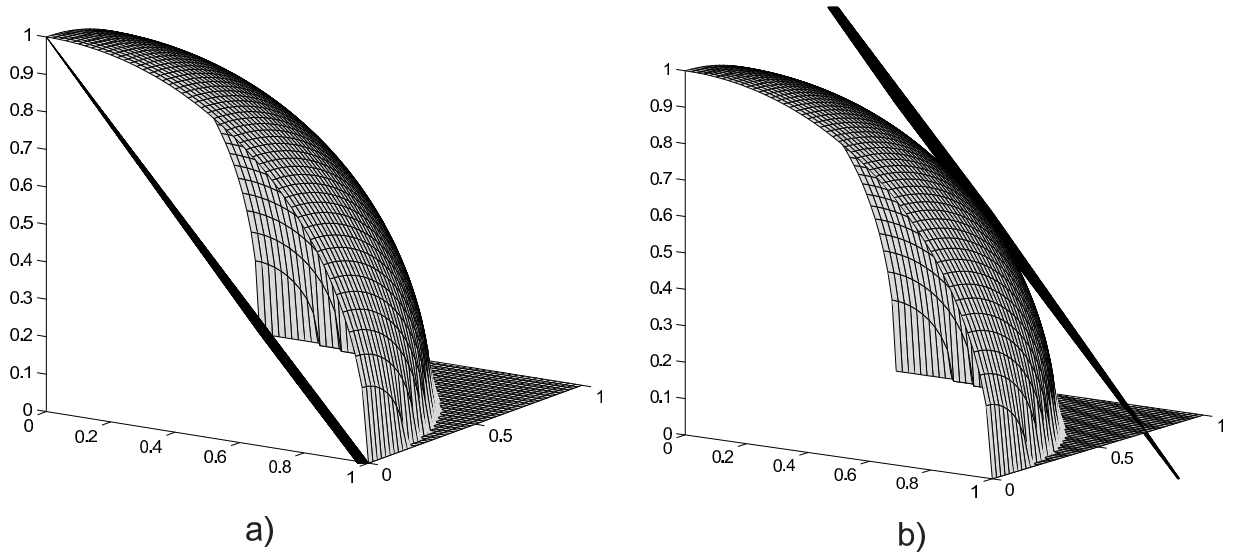}}
\caption{\footnotesize The plane and the sphere are intersected when the plane is placed between the two limiting positions shown on the pictures a) and b). The plane on the figure b) is tangent to the sphere.}
\label{sphere.and.plane}
\end{figure}

Obviously, for the system~(\ref{h2h4}) to have  nontrivial solutions it is necessary that
\eq{\omega-\dac{1}{3}<0,\quad9\(\omega-\dac{1}{3}\)^2\(\dac{\varkappa}{2}\)^2+\dac{\omega-1}{\alpha}\(\dac{\varkappa}{2}\)>0}
We see, first of all, that the equation of state parameter $w$ is restricted from above: $w<1/3$. However, positivity of quadratic and quartic sums is not sufficient for the solution to exist. Going further we denote:
\eq{\xi_{1}=\frac{\varkappa}{2},\quad\xi_{2}=\frac{1}{\alpha}\(\frac{\varkappa}{2}\),\quad\xi=\frac{|\xi_2|}{\xi_1^2}}
\eq{a=\xi_1(1-3\omega),\quad r^2=\frac{1}{2}\[\xi_1^2(1-3\omega)^2+\xi_2(\omega-1)\],\quad\eta_k=H_k^2}
Then equations~(\ref{h2h4}) take the form:
\eq{\sum\limits_{i}\eta_i=a,\quad\sum\limits_{i}\eta_i^2=r^2\label{plane.and.sphere}}
Variables $\eta_1,\ldots,\eta_N$ can be considered as Cartesian coordinates in $N$-dimensional Euclidean space; then the first of the equations~(\ref{plane.and.sphere}) specifies $(N-1)$-dimensional hyperplane which intersects each axis of the coordinate system at the point $a$, the second of the equations~(\ref{plane.and.sphere}) describes $(N-1)$-dimensional hypersphere of radius $r$ centred at the origin. Since $a>0$ and all $\eta_i>0$ we deal with fragments of the hypersphere and the hyperplane located in the first orthant. These fragments are intersected iff $r\leqslant a\leqslant\sqrt{N}r$.
Fig.~\ref{sphere.and.plane} illustrates this reasoning for the 3D case.
Obviously,
\eq{\left\{\begin{array}{l}
             r\leqslant a\leqslant\sqrt{N}r \\
             a>0
           \end{array}\right.
           \iff
           \frac{1}{N}\leqslant\frac{r^2}{a^2}\leqslant1
\label{r<a<Nr}}
So, system~(\ref{h2h4}) has nontrivial solutions iff $\frac{r^2}{a^2}\in\left[\frac{1}{N};1\right]$. We are concerned in such solutions of the system~(\ref{h2h4}) that satisfy the condition $\sum_iH_i=0$. It turns out that there is essential difference between even- and odd-dimensional cases. Indeed, let us consider $(4+1)$-dimensional spacetime; Eqs.~(\ref{plane.and.sphere}) describe 4-plane and 4-sphere; in the point of contact of these surfaces we have $H_1^2=H_2^2=H_3^2=H_4^2$, therefore, one can choose $H_1,\ldots,H_4$ such that $H_1=H_2=-H_3=-H_4$ and the condition $H_1+\ldots+H_4=0$ is satisfied automatically. Clearly, there is no way one can satisfy the condition $H_1+\ldots+H_5=0$ in the point of tangency of 5-plane and 5-sphere because of one extra positive (or negative) summand. This results can be generalized to the case of arbitrary dimension: for an even-dimensional spacetime there exist solutions of the equations $\sum\limits_{i}H_i^2=a,\quad\sum\limits_{i}H_i^4=r^2$ such that $\sum_iH_i=0$ in the vicinity of the point of contact hyperplane and hypersphere specified by Eqs.~(\ref{plane.and.sphere}); for an odd-dimensional spacetime there are no solutions in the vicinity the aforementioned point of contact. In general, there exist a subset $I\subseteq\left[\frac{1}{N};1\right]$ such that
\eq{\sum\limits_{i}H_i=0,\quad\sum\limits_{i}H_i^2=a,\quad \sum\limits_{i=2}^NH_i^4=r^2\quad\mbox{for all}\quad\frac{r^2}{a^2}\in I\label{basic.system}}
We express one of the Hubble parameters from the first of Eqs.~(\ref{basic.system}) and substitute it in the remaining equations:
\eq{\(\sum\limits_{i=2}^NH_i\)^2+\sum\limits_{i=2}^NH_i^2=a,\quad\(\sum\limits_{i=2}^NH_i\)^4+\sum\limits_{i}H_i^4=r^2\label{basic.system.after.subs}}
Hubble parameters can be considered here as a Cartesian coordinates; after reducing the quadratic form $\(\sum\limits_{i=2}^NH_i\)^2+\sum\limits_{i=2}^NH_i^2$ to the canonical form by a coordinate transformation and converting a Cartesian coordinates to spherical ($\varrho,\theta_1,\ldots,\theta_{N-2}$) Eqs.~(\ref{basic.system.after.subs}) take the form correspondingly:
\eq{\varrho^2=a,\quad\varrho^4f(\theta_1,\ldots,\theta_{N-2})=r^2\,,\label{basic.system.after.subs.sph.coord}}
where $f$ is a polynomial in $\sin(\theta_k),\cos(\theta_k)\;\mbox{for}\;k=\overline{1,N-2}$. Substituting $\varrho^2=a$ into the second of Eqs.~(\ref{basic.system.after.subs.sph.coord}) we obtain
\eq{F(\theta_1,\ldots,\theta_{N-2},r,a)=0,\quad F(\theta_1,\ldots,\theta_{N-2},r,a)=f(\theta_1,\ldots,\theta_{N-2})-\frac{r^2}{a^2}}
For example, for $N=4$ we have:
\eq{\begin{array}{c}
      F(\theta_1,\theta_{2},r,a)=\frac{1}{4}\sin^4\left(\theta_{{1}}\right)    \left(\frac{1}{2}+\frac{11}{6}\cos^4\left(2\theta_{{2}}\right)\right)+\frac{1}{2}\Bgp{\cos^{2}\left(     \theta_{{1}}\right)+\sin^{2}\left(\theta_{{1}}\right)\sin^{2}\left(2\theta_{{2}}\right)}^{2}+\\
      +\frac{1}{8}\sin^2\left(2\theta_{{1}}\right)\cos^2\left(2\,\theta_{{2}}\right)+
      \frac{1}{3\sqrt{2}}\sin\left(2\theta_{{1}}\right)\cos\left(2\theta_{{2}}\right)\Bgp{\cos^{2}\left(\theta_{{1}}\right)-3\sin^2\left(\theta_{{1}} \right)\sin^2\left(2\,\theta_{{2}}\right)}-\frac{r^2}{a^2}
    \end{array}
}
So, the problem of the existence of solutions of Eqs.~(\ref{basic.system}) is reduced to the problem of the existence of zeros of function $F$. We solve this problem numerically. Numerical calculations performed for $N\in\bgbr{4,\ldots,8}$ shows that functions $F$ has zeros for $\frac{r^2}{a^2}\in[\sigma_+;\sigma_-]$, i.e $I=[\sigma_+;\sigma_-]$ such that $\frac{1}{N}\leqslant\sigma_+<\frac{1}{2}<\sigma_-<1$.
Using this fact we get:
\eq{\sigma_+\leqslant\frac{r^2}{a^2}\leqslant\sigma_-\iff
\left\{\begin{array}{c}
         \bgsb{2\sigma_+-1}\xi_1^2(1-3\omega)^2\leqslant\xi_2(\omega-1) \\
         \bgsb{2\sigma_--1}\xi_1^2(1-3\omega)^2\geqslant\xi_2(\omega-1)
       \end{array}\right.
}
There are two cases, depending on the sign of the parameter $\alpha$.

\textbf{I.} $\alpha>0$.
\eq{\left\{\begin{array}{c}
         \bgsb{2\sigma_+-1}\xi_1^2(1-3\omega)^2\leqslant\xi_2(\omega-1) \\
         \bgsb{2\sigma_--1}\xi_1^2(1-3\omega)^2\geqslant\xi_2(\omega-1)
       \end{array}\right.
       \iff
       \omega\leqslant\frac{1}{3}-\frac{\xi_++\sqrt{\xi_+(\xi_++24)}}{18},\quad\xi_+=\frac{\xi}{|2\sigma_+-1|}
\label{c2>0-2}}

\textbf{II.} $\alpha<0$.
\eq{\left\{\begin{array}{c}
         \bgsb{2\sigma_+-1}\xi_1^2(1-3\omega)^2\leqslant\xi_2(\omega-1) \\
         \bgsb{2\sigma_--1}\xi_1^2(1-3\omega)^2\geqslant\xi_2(\omega-1)
       \end{array}\right.
       \iff
       \omega\leqslant\frac{1}{3}-\frac{\xi_-+\sqrt{\xi_-(\xi_-+24)}}{18},\quad\xi_-=\frac{\xi}{2\sigma_--1}
\label{c2<0-2}}
Finally we obtain:
\eq{\left\{\begin{array}{l}
             \sum\limits_{i}H_i=0 \vspace{.2cm}\\
             \sum\limits_{i}H_i^2=-3\(\omega-\dac{1}{3}\)\(\dac{\varkappa}{2}\) \vspace{.2cm}\\
             \sum\limits_{i}H_i^4=\dac{1}{2}\left[9\(\omega-\dac{1}{3}\)^2\(\dac{\varkappa}{2}\)^2+\dac{\omega-1}{\alpha}\(\dac{\varkappa}{2}\)\right]
           \end{array}\right.
\iff\omega\leqslant
\left\{\begin{array}{l}
         \dac{1}{3}-\dac{\xi_++\sqrt{\xi_+(\xi_++24)}}{18},\;\alpha>0 \vspace{.2cm}\\
         \dac{1}{3}-\dac{\xi_-+\sqrt{\xi_-(\xi_-+24)}}{18},\;\alpha<0
       \end{array}\right.
\label{full.sys.solution}}
Inequalities~(\ref{full.sys.solution}) can be rewritten in terms of the energy density $\rho$:
\eq{\rho\geqslant\rho_{\rm lim}(\omega),\quad\rho_{\rm lim}(\omega)=\left\{\begin{array}{l}
         \dac{1}{36\pi\alpha}\dac{1}{2\sigma_+-1}\dac{\omega-1}{\(\omega-\frac{1}{3}\)^2},\;\alpha>0 \vspace{.2cm}\\
         \dac{1}{36\pi\alpha}\dac{1}{2\sigma_--1}\dac{\omega-1}{\(\omega-\frac{1}{3}\)^2},\;\alpha<0
       \end{array}\right.\label{rho}}
We see that the above mentioned nonexistence of vacuum solutions has a sharper form: for any $\omega$ there exists a low limit for $\rho$. In the particular case of cosmological constant $\omega=-1$:
\eq{\rho_{\rm lim}(-1)=\left\{\begin{array}{l}
         -\dac{1}{32\pi\alpha}\dac{1}{2\sigma_+-1},\;\alpha>0 \vspace{.2cm}\\
         -\dac{1}{32\pi\alpha}\dac{1}{2\sigma_--1},\;\alpha<0
       \end{array}\right.}
Since the function $\rho_{\rm lim}(\omega)$ is growing, for any non-fantom ($\omega\geqslant-1$) matter we have $\rho\geqslant\rho_{\rm lim}(-1)$.
Let us discuss briefly the problem of finding of the parameters $\sigma_+,\sigma_-$. We consider the cases of an even-dimensional and an odd-dimensional space separately.

\textbf{I.} It is easy to show that $\sigma_+=\frac{1}{N}$ when the number $N$ of space dimensions is even. Indeed, let $N$ be an even number, in this case one can satisfy the condition $\sum_{i}H_i=0$ by choosing parameters $H_1,\ldots,H_N$ such that
\eq{H_1=-H_2,\ldots,H_{N-1}=-H_{N}\label{H1=-H2}}
It follows from~(\ref{H1=-H2}) that $\eta_1=\eta_2,\ldots,\eta_{N-1}=\eta_N$, and equations~(\ref{plane.and.sphere}) take the form:
\eq{\sum\limits_{i=1}^{N/2}\eta_i=\frac{a}{2},\quad \sum\limits_{i=1}^{N/2}\eta_i^2=\frac{r^2}{2},\quad\eta_i>0\label{plane.and.sphere.for.N/2}}
Repeating the above arguments we deduce that system~(\ref{plane.and.sphere.for.N/2}) has nontrivial solutions iff
\eq{\frac{r}{\sqrt{2}}\leqslant \frac{a}{2}\leqslant\frac{r}{\sqrt{2}}\sqrt{\frac{N}{2}}\iff\frac{1}{N}\leqslant\frac{r^2}{a^2}\leqslant\frac{1}{2}\label{r/2<a/2<Nr/2}}
We see that $\sigma_+=\frac{1}{N}$; substitution it into~(\ref{rho}) leads to
\eq{\rho\geqslant\rho_{\rm lim}(\omega)=-\dac{1}{36\pi\alpha}\dac{N}{N-2}\dac{\omega-1}{\(\omega-\frac{1}{3}\)^2},\;\alpha>0\label{rho.lim.N.even}}
The case $\rho=\rho_{\rm lim}$ corresponds exactly to the situation when plane $\sum\limits_{i=1}^{N}\eta_i=a$ touches sphere $\sum\limits_{i=1}^{N}\eta_i^2=r^2$ and $\eta_1=\eta_2,\ldots,\eta_{N-1}=\eta_N$; in view of the condition $\sum_{i}H_i=0$ the latter implies~(\ref{H1=-H2}). In the previous section we described this case as $\(\frac{N}{2}+\frac{N}{2}\)$-decomposition; it is easy to check that $\rho_{\rm lim}$ in~(\ref{rho.lim.N.even}) matches with $\rho$ given in~(\ref{sol.for.N/2+N/2}).

The upper threshold $\sigma_-$ is a bit large than $\frac{1}{2}$ (condition~(\ref{H1=-H2}) is not necessary: there exist other sets of the Hubble parameters such that $\sum_iH_i=0$). For $N=4$ numerical calculations give $\sigma_-=0.76\pm0.01$. So, for $(4+1)$-dimensional spacetime we have:
\eq{\label{last1}\omega<
\left\{\begin{array}{l}\
         \dac{1}{3}-\dac{2\xi+\sqrt{2\xi+(2\xi+24)}}{18},\;\alpha>0 \\
         \dac{1}{3}-\dac{1.92\xi+\sqrt{1.92\xi(1.92\xi+24)}}{18},\;\alpha<0
       \end{array}\right.
       \,,\quad\mbox{or}\quad
       \rho\gtrsim\left\{\begin{array}{l}
         -\dac{1}{18\pi\alpha}\dac{\omega-1}{\(\omega-\frac{1}{3}\)^2},\;\alpha>0 \vspace{.2cm}\\
         -\dac{1}{1.04}\dac{1}{18\pi\alpha}\dac{\omega-1}{\(\omega-\frac{1}{3}\)^2},\;\alpha<0
       \end{array}\right.
}
where $\xi=\frac{1}{4\pi|\alpha|\rho}$.

\textbf{II.} When working with an odd-dimensional space we have no possibility to satisfy the condition $\sum_{i}H_i=0$ for $\sigma_+=\frac{1}{N}$. Indeed, system~(\ref{plane.and.sphere}) has the only solution $\eta_1=\ldots=\eta_N=\frac{r}{\sqrt{N}}$ for $\frac{r^2}{a^2}=\frac{1}{N}$ (geometrically it corresponds to point of contact plane and sphere -- see Fig.~\ref{sphere.and.plane}a). Since $H_1=\ldots=H_N=\pm\sqrt{\frac{r}{\sqrt{N}}}$, the sum $\sum_{i}H_i$ has at least one extra positive (or negative) term and can not vanish for odd $N$. For $N=5$ numerical calculations give $\sigma_+=0.23\pm0.01,\;\sigma_-=0.65\pm0.01$. So, for $(5+1)$-dimensional spacetime we have:
\eq{\label{last2}\omega<
\left\{\begin{array}{l}
         \dac{1}{3}-\dac{1.85\xi+\sqrt{1.85\xi+(1.85\xi+24)}}{18},\;\alpha>0 \\
         \dac{1}{3}-\dac{3.33\xi+\sqrt{3.33\xi(3.33\xi+24)}}{18},\;\alpha<0
       \end{array}\right.
       \,,\quad\mbox{or}\quad
       \rho\gtrsim\left\{\begin{array}{l}
         \dac{1}{0.54}\dac{1}{36\pi\alpha}\dac{\omega-1}{\(\omega-\frac{1}{3}\)^2},\;\alpha>0 \vspace{.2cm}\\
         -\dac{1}{0.3}\dac{1}{36\pi\alpha}\dac{\omega-1}{\(\omega-\frac{1}{3}\)^2},\;\alpha<0
         \end{array}\right.
}
where $\xi=\frac{1}{4\pi|\alpha|\rho}$.

\section {Conclusions}

In the present paper we have considered solutions with constant different Hubble parameters
in a flat Einstein-Gauss-Bonnet cosmology. Such solutions are absent in a pure Einstein
gravity and its existence represent one of specific features of higher-order curvature terms (see~\cite{PT} for details).

Let us discuss the results of this paper alone first, and then draw the conclusions for this paper combined with results of~\cite{CST}, for which current paper
could be considered as a direct continuation.

So in this paper we generalize some of the previously obtained solutions on the case with only two distinct Hubble parameters. As it was mentioned in the Introduction,
with increase of the number of dimensions and the order of Lovelock correction, the complexity of the equations increase drastically, making the quest for finding
exact solutions more challenging. But there are cases when we still can find them almost regardless of the dimensionality and the order of the corrections, and one
of these cases is the case with only two distinct Hubble parameters. Despite the fact that it looks unnatural, it still holds its meaning representing, say, the case when
there are two manifolds with two different scale factors in the spatial section. Of the special interest is the case with one of these spatial sections to be three-dimensional
since it could represent our three-dimensional (spatially) Universe (see~\cite{CST2}). Also, generalized solutions could exist in presence of perfect fluid with
$\omega < 1/3$ unlike their original counterparts~\cite{CST} which could exist only for vacuum/$\Lambda$-term cases.

The second important result is the complete description of the constant-volume solutions. The importance of these solutions lies in the fact that they are allowed for
a wide range of equations of state of the perfect fluid (see (\ref{last1}) and (\ref{last2})). Indeed, as it could be seen from (\ref{rho_all}), solutions with
varying volume could be obtained only in vacuum ($\rho\equiv 0$) or $\Lambda$-term ($\omega=-1$) cases. So this is the only class of exponential solutions which could be
obtained in the presence of perfect fluid with a range of equations of state.

Combining results of the present paper with results of our previous paper~\cite{CST} we can
write down full classification of solutions in question in (4+1) and (5+1) dimensions.

\begin{itemize}

\item {Vacuum solution in a pure Gauss-Bonnet gravity \cite{Ivashchuk}. We have shown
 that this solution is a particular one and can not be incorporated in other sets of
solution of the type considered. It requires absence of both matter and Einstein-Hilbert term.}

\item {Solutions with volume element changing in time. Such solutions require a matter only
in the form of cosmological constant. Apart from an
isotropic solution, it appears that these solutions
exist only when set of Hubble parameters is divided into subsets with equal values of Hubble parameters
belonging to
the same subset (so, existence of isotropic subspaces is required).  }

\item {Solutions with constant volume element. Solutions of this type exist only when matter density exceeds (or equal to) some critical
value which depends on the equation of state of the matter. The parameter $\omega$ of the matter should be smaller than $1/3$. In general,
solutions do not have isotropic subspaces, though can have them for special cases.}

\end{itemize}

As space-times with isotropic subspaces represent a particular interest (for example, if
multidimensional paradigm is indeed realized in Nature, then our own world belongs to
this class) we write down explicit solutions of constant volume element with isotropic subspaces,
generalising those found in \cite{CST}. For a general case of constant volume
element (without isotropic subspaces) we present the  conditions for such solutions to exist,
leaving their explicit form  to a future work.

\textit{Acknowledgments.-- }
This work was supported by RFBR grant No. 14-02-00894. S.A.P. was supported by FONDECYT via grant No. 3130599.
Authors are grateful to Vitaly Melnikov and Vladimir Ivashchuk for discussions.

\end{document}